\documentclass[aps,pra,twocolumn,nofootinbib]{revtex4-1}
\usepackage{epsfig}
\usepackage[colorlinks,linkcolor=blue,anchorcolor=blue,citecolor=blue,urlcolor=blue,breaklinks=true]{hyperref}
\usepackage{graphics}
\usepackage{color}
\usepackage{amsmath}
\usepackage{amsfonts}
\usepackage{bm}
\begin{document}
\author{Guo-Hua Liang$^{1}$}
\email{lianggh@njupt.edu.cn}
\author{Ai-Guo Mei$^{1}$}
\author{Men-Yun Lai$^{2}$}
\author{Shu-Sheng Xu$^{1}$}
\affiliation{$^{1}$ School of Science, Nanjing University of Posts and Telecommunications, Nanjing 210023, China}
\affiliation{$^{2}$ College of Physics and Communication Electronics, Jiangxi Normal University, Nanchang 330022, China}

\title{Quantum States in Twisted Tubes with Linear Cross-Section Variation}
\begin{abstract}
We study the quantum dynamics of a particle confined in a twisted tube with a linearly varying cross section. By relating a general linear transformation matrix to the system's Hamiltonian, we use an extended thin-layer method to derive an effective Hamiltonian for tangential motion under mild and general linear transformations. Explicit forms are provided for three fundamental transformations: rotation, scaling, and shearing. Rotation introduces a gauge field coupled to angular momentum, while scaling and shearing produce geometric potentials that lift degeneracies in non-circular cross sections. In square cross sections, these transformations cause energy splittings among formerly degenerate states, whereas circular cross sections retain degeneracy. Through an example combining rotation and squeezing, we analyze state evolution and compute the quantum geometric tensor to quantify geometric response. Our results demonstrate how geometric transformations can tailor quantum states and suggest that circular waveguides are more robust against mode mixing.
\end{abstract}


\maketitle
\section{INTRODUCTION}
Quantum particles and photons confined in nanoscale systems can exhibit behavior similar to that observed in the presence of strong background fields~\cite{vsvanvcara2024rotating,Bekenstein2017Control}. This occurs because the characteristic length scales in such systems approach the particle's wavelength, making nanoscale confinement an excellent platform for exploring a wider range of fundamental phenomena. In recent decades, advances in nanostructure fabrication~\cite{Pogosov_2022,lilian2022curvature,Gentile2022,nano12183226,wu2025step} have enabled laboratory studies of various nanoscale dynamics. For example, scaled vertical-nanowire heterojunction tunnelling transistors are realized based on extreme quantum confinement in GaSb/InAs system~\cite{shao2025scaled}, a room-temperature nonlinear Hall effect has been observed in polycrystalline bismuth thin films~\cite{makushko2024tunable}, and the energy landscape of Bloch points in ferromagnetic nanowires can be tailored by curvature~\cite{ruiz2025tailoring}.

Certain nanostructures, especially low-dimensional ones, exhibit significant curvature and torsion. The effects of these geometric quantities have been investigated theoretically in various 2D and 1D systems. Using the thin-layer approach, it has been found that in 2D systems, curvature induces a scalar quantum potential known as the geometric potential~\cite{JENSEN1971586,PhysRevA.23.1982,PhysRevLett.104.150403,PhysRevB.84.045438,Wang2017}. When the particle possesses internal degrees of freedom~\cite{PhysRevA.48.1861,BRANDT20163036,PhysRevB.64.085330,PhysRevB.87.174413, PhysRevA.98.062112,PhysRevA.78.043821} or the system is subjected to external fields ~\cite{PhysRevLett.100.230403,PhysRevA.90.042117}, curvature can act as an effective electric or magnetic field. In addition to curvature effects, torsion in quasi-1D and higher-dimensional systems plays the role of an induced gauge field that couples to the topological charge in the degenerate state space~\cite{OUYANG1999297,PhysRevB.91.245412,PhysRevA.97.033843,PhysRevA.100.033825,S0217732393000891,Maraner_1995,MARANER1996325,S0217751X97002814,SCHUSTER2003132}. This feature in closed 1D systems is related to the Berry phase~\cite{ptp/87.3.561} and gives rise to a geometrically induced Aharonov-Bohm effect~\cite{sciadv.aau8135}.

Most of these theoretical work adopts the idealized assumption that particles are confined to smooth surface layers or curved tubes with constant thickness or radius. However, defects are inevitable during the synthesis of nanomembranes or nanowires. By extending the thin-layer approach, it has been found that a slightly inhomogeneous thickness in a curved layer can induce an effective potential in the tangential dynamics, proportional to the ground-state energy and the thickness morphology function~\cite{PhysRevA.107.022213}. This potential allows quantum states to be localized in specific regions of the curved surface through engineered thickness distributions~\cite{LIANG2025170144}. Further study indicates that thickness fluctuations do not affect spin dynamics~\cite{Liang_2024}.

In this work, we employ the extended thin-layer approach to investigate the quantum dynamics of a particle confined within a twisted tube with a cross section that varies linearly along its axis. Studying quantum dynamics in tubes with varying cross sections is more challenging than in surface layers with nonuniform thickness, as the changing constrained dimension introduces additional complexity. A linear transformation of the two-dimensional cross section can be decomposed into fundamental transformations, namely rotation, scaling, and shearing~\cite{farin2021practical}. We formulate the effective quantum dynamics in tubes under general linear transformations and examine the effects of each type of fundamental transformation individually. Furthermore, we investigate the sensitivity of quantum states, particularly degenerate states, to changes in transformation parameters, as captured by the quantum geometric tensor~\cite{cheng2010quantum,PhysRevB.81.245129,gianfrate2020measurement,kang2025measurements}. The real part of this tensor defines a Riemannian metric on the parameter space, providing a measure of the distance between quantum states, while the imaginary part corresponds to the Berry curvature, which governs the geometric phase of the system.

This paper is structured as follows. Section~\ref{sec2} builds the relationship between a general linear transformation matrix to the system Hamiltonian. Section~\ref{sec3} applies the condition of slight transformation and presents the explicit form of the effective Hamiltonian for rotation, scaling, and shearing of the cross section, respectively. In Section~\ref{sec4}, we discuss the energy splitting, states phase and quantum geometric tensor in an example where a twisted tube experiences rotation and squeezing simultaneously. Finally, our conclusion are written in Section~\ref{sec5}.

\section{Hamiltonian in adapted coordinates}\label{sec2}
\begin{figure}
  \centering
  \includegraphics[width=0.45\textwidth]{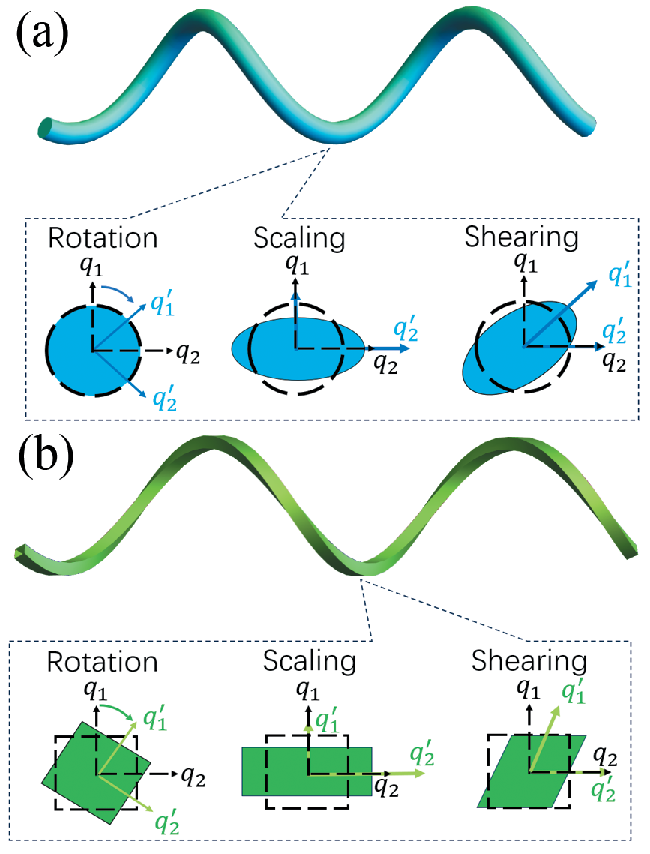}
  \caption{(a) Schematic of a twisted tube with a circular cross section undergoing linear transformations: rotation, scaling, and shearing.
(b) Same as (a), but with a square cross section.}\label{fig1}
\end{figure}

We consider a quantum particle confined to a twisted tube with its axis forming a space curve $\mathcal{C}$, as illustrated in Fig.~\ref{fig1}. The cross section of the tube varies linearly along $\mathcal{C}$. In Fig.~\ref{fig1}, we demonstrate three fundamental linear transformations, namely rotation, scaling and shearing, which are applied to both circular and square cross sections. The $q_1$-$q_2$ plane represents the normal plane of $\mathcal{C}$, where $q_1$ and $q_2$ are coordinates along the normal vector $\bm{n}$ and the binormal vector $\bm{b}$, respectively. The tangent vector $\bm{t}$ of $\mathcal{C}$, together with $\bm{n}$ and $\bm{b}$, forms the Frenet frame. In this frame, the neighborhood space around $\mathcal{C}$ can be parameterized as
\begin{equation}
\bm{R}=\bm{r}(s)+q_1\bm{n}(s)+q_2\bm{b}(s),
\end{equation}
where $\bm{r}(s)$ is the position vector of $\mathcal{C}$ and $s$ represents the arclength. The Frenet frame satisfies the Frenet-Serret equation
\begin{equation}\label{fse}
\left(
\begin{array}{ccc}
\dot{\bm{\mathfrak{t}}}\\
\dot{\bm{n}}\\
\dot{\bm{b}}
\end{array}
\right)=\left(
\begin{array}{ccc}
0&\kappa(s)&0\\
-\kappa(s)&0&\tau(s)\\
0&-\tau(s)&0
\end{array}
\right)\left(
\begin{array}{ccc}
\bm{\mathfrak{t}}\\
\bm{n}\\
\bm{b}
\end{array}
\right),
\end{equation}
where the dot denotes the derivative with respect to $s$; $\kappa(s)$ and $\tau(s)$ are the curvature and torsion of the curve, respectively.

To accommodate the variation in the cross section of the tube, we introduce new coordinates $(q_1', q_2')$ in the normal plane of $\mathcal{C}$, which are related to the original coordinates $(q_1, q_2)$ via a $2 \times 2$ linear transformation matrix $\mathbb{T}(s)$, namely
\begin{equation}
\left(
\begin{array}{ccc}
q_1 \\ q_2
\end{array}
\right)=\mathbb{T}(s) \left(
\begin{array}{ccc}
q_1' \\ q_2'
\end{array}
\right).
\end{equation}
Here, the components $T_{ab}(s)$, with $a, b = 1, 2$, are continuous functions of $s$. The matrix can be expressed as $\mathbb{T} = (\bar{\bm{t}}_1, \bar{\bm{t}}_2)^\mathrm{T} = (\tilde{\bm{t}}_1, \tilde{\bm{t}}_2)$, where $\bar{\bm{t}}_a = (T_{a1}, T_{a2})$, $\tilde{\bm{t}}_a = (T_{1a}, T_{2a})^\mathrm{T}$, and the superscript $\mathrm{T}$ denotes the transpose.

In the new coordinate system, the space around $\mathcal{C}$ can be parameterized as
\begin{equation}
\begin{aligned}
\bm{R}=\bm{r}(s)+(\bar{\bm{t}}_1\cdot\bm{q}'_{\perp})\bm{n}(s) +(\bar{\bm{t}}_2\cdot\bm{q}'_{\perp})\bm{b}(s),
\end{aligned}
\end{equation}
where $\bm{q}'_{\perp} = (q'_1, q'_2)^\mathrm{T}$.
A straightforward calculation yields
\begin{equation}
\partial_{q_1}\bm{R}=T_{11}\bm{n}+T_{21}\bm{b},
\end{equation}
\begin{equation}
\partial_{q_2}\bm{R}=T_{12}\bm{n}+T_{22}\bm{b},
\end{equation}
and
\begin{equation}
\begin{aligned}
\partial_s\bm{R}
=\gamma\bm{\alpha}+\zeta\bm{n} +\eta\bm{b},
\end{aligned}
\end{equation}
where $\hat{\textbf{d}} = (\partial_s, \tau)^\mathrm{T}$ is a vector operator, $\gamma = 1 - \kappa (\bar{\bm{t}}_1 \cdot \bm{q}'_{\perp})$, $\zeta = (\sigma_z \hat{\textbf{d}})^\mathrm{T} \mathbb{T} \bm{q}'_{\perp}$, and $\eta = (\sigma_x \hat{\textbf{d}})^\mathrm{T} \mathbb{T} \bm{q}'_{\perp}$. Here, $\sigma_x$ and $\sigma_z$ denote the Pauli matrices.

In the curvilinear coordinates $(q_1, q_2, s)$, the metric tensor can be obtained through $G_{ij} = \partial_i \bm{R} \cdot \partial_j \bm{R}$, where $i, j = 1, 2, 3$. The explicit form of the metric is given by the matrix
\begin{equation}
G_{ij}=\left(
\begin{array}{ccc}
|\tilde{\bm{t}}_1|^2 & \tilde{\bm{t}}_1\cdot\tilde{\bm{t}}_2 & \bm{\xi}\cdot \tilde{\bm{t}}_1 \\
\tilde{\bm{t}}_1\cdot\tilde{\bm{t}}_2 & |\tilde{\bm{t}}_2|^2 & \bm{\xi}\cdot \tilde{\bm{t}}_2 \\
\bm{\xi}\cdot \tilde{\bm{t}}_1& \bm{\xi}\cdot \tilde{\bm{t}}_2& \gamma^2+\zeta^2+\eta^2
\end{array}
\right),
\end{equation}
where $\bm{\xi} = (\zeta, \eta)^\mathrm{T}$.
The determinant of the metric is then $G=\det(G_{ij})=(1-\kappa \bar{\bm{t}}_1\cdot\textbf{q}'_{\perp})^2 |\mathbb{T}|^2$.

For the inverse of the metric, the components can be expressed in compact form as follows
\begin{equation}
G^{ss}=\frac{1}{\gamma^2},
\end{equation}
\begin{equation}
\begin{aligned}
G^{sa}=G^{as}=\frac{\epsilon^{ac}\bm{t}_c \cdot \bar{\bm{\xi}}}{\gamma^2|\mathbb{T}|},
\end{aligned}
\end{equation}
and
\begin{equation}
G^{ab}=\gamma^2 G^{sa}G^{sb}+\frac{(\epsilon^{ac}\tilde{\textbf{t}}_c) \cdot (\epsilon^{bd}\tilde{\textbf{t}}_d)}{|\mathbb{T}|^2},
\end{equation}
where $\epsilon^{ac}$ denotes the Levi-Civita symbol and $\bar{\bm{\xi}}=i\sigma_y \bm{\xi}$.

Now let us discuss the influence of cross section shape variations on the effective dynamics. We assume that the particles are confined to a twisted tube by a confining potential $V_c$, which has a deep minimum at $q_1 = q_2 = 0$. If the cross section remains constant along the axis (as indicated by the dashed outlines in Fig.~\ref{fig1}), i.e., $V_c = V_c(q_1, q_2)$, the bound energy levels in the transverse direction are maintained. This facilitates a separation of dynamics between the longitudinal and transverse directions. However, when the cross section varies along the axis, $V_c$ becomes $s$-dependent, which prevents such a separation. We expect that the coordinate transformation $\mathbb{T}$, adapted to the cross section variation, will restore the $s$-independence of the confining potential, transforming $V_c(q_1, q_2, s)$ into $V'_c(q'_1, q'_2)$.

In the curvilinear coordinates $(q_1', q_2', s)$, the Schr\"{o}dinger equation for a particle confined within the tube is given by
\begin{equation}
-\frac{\hbar^2}{2m}\nabla^2 \Psi+V_c'(q_1',q_2')\Psi=E\Psi,
\end{equation}
where the Laplacian operator corresponding to the given metric takes the form
\begin{equation}
\begin{aligned}
\nabla^2=&\frac{1}{\sqrt{G}}\partial_i(\sqrt{G}G^{ij}\partial_j)  \\
=&\frac{1}{\sqrt{G}}\frac{(\epsilon^{ac}\tilde{\bm{t}}_c) \cdot (\epsilon^{bd}\tilde{\bm{t}}_d)}{|\mathbb{T}|^2} \partial_a \sqrt{G} \partial_b+\frac{1}{\sqrt{G}}\partial_s(G^{ss}{\sqrt{G}}\partial_s) \\ &+\frac{1}{\sqrt{G}}\partial_a(\sqrt{G}G^{ss} \frac{\mathcal{A}^a\mathcal{A}^b}{|\mathbb{T}|^2}\partial_b) \\
&+ \frac{1}{\sqrt{G}}\partial_s(\sqrt{G}G^{sa}\partial_a)+
\frac{1}{\sqrt{G}}\partial_a(\sqrt{G}G^{sa}\partial_s),
\end{aligned}
\end{equation}
with the definition $\mathcal{A}^a = \epsilon^{ac} \tilde{\bm{t}}_c \cdot \bar{\bm{\xi}}$.
By introducing a derivative operator $D_s=\partial_s+\frac{1}{2}\frac{\{\mathcal{A}^a,\partial_a\}}{|\mathbb{T}|}$, where $\{\cdot,\cdot\}$ denotes the anticommutator, we can compactly rewrite the Laplacian as
\begin{equation}
\nabla^2=\frac{1}{\sqrt{G}}D_s \sqrt{G}G^{ss}D_s+\frac{1}{\sqrt{G}}\frac{\epsilon^{ac}\epsilon^{bd}\tilde{\bm{t}}_c \cdot \tilde{\bm{t}}_d}{|\mathbb{T}|^2} \partial_a \sqrt{G} \partial_b-V_T
\end{equation}
where
\begin{equation}
\begin{aligned}
V_T=&\frac{1}{2\gamma|\mathbb{T}|^2}\mathcal{A}^a\partial_a\left[\frac{1}{\gamma} (\partial_b \mathcal{A}^b )\right] +\frac{1}{4\gamma^2|\mathbb{T}|^2}(\partial_a \mathcal{A}^a) (\partial_b \mathcal{A}^b) \\
&+\frac{1}{2\gamma|\mathbb{T}|^2}\partial_s\left[\frac{|\mathbb{T}|}{\gamma}\partial_a \mathcal{A}^a\right].
\end{aligned}
\end{equation}
Here we establish the relationship between the cross-sectional transformation matrix and the equation of motion. It should be noted that when the transformation matrix reduces to the identity matrix, $\mathcal{A}^a$ simplifies to $\epsilon^{ac} \tau q_c$, and the anticommutator in the covariant derivative reduces to $-\frac{i}{\hbar} \tau \hat{L}$~\cite{SCHUSTER2003132,ptp/87.3.561}, where the angular momentum operator $\hat{L} = i\hbar (q_2 \partial_{q_1} - q_1 \partial_{q_2})$.

\section{Thin-layer procedure for slight transformations}\label{sec3}
To obtain the tangential dynamics under the confining potential, it is necessary to distinguish between different energy regimes. For this purpose, we introduce a dimensionless small parameter $\delta$. In the energy regime under consideration, the wave function is localized near $q_1 = q_2 = 0$. We therefore rescale the transverse coordinates as $\bm{q}_\perp '\rightarrow \sqrt{\delta}\bm{q}_\perp '$. To match the kinetic energy in the transverse direction, the confining potential should be rescaled as $V_c'\rightarrow V_c'/\delta$. Considering the metric in the adapted coordinates, the wave function satisfies the normalization condition $\int |\Psi|^2 \sqrt{G}ds d\bm{q}_\perp '=1$. By introducing a new wave function $\psi = G^{1/4} \Psi$, we obtain a one-dimensional probability density along the axial direction defined as $P(s)=\int |\psi|^2 d\bm{q}_\perp '$, which satisfies the normalization condition $\int P(s) ds=1 $.
The Hamiltonian corresponding to the new wave function is then transformed as $H\rightarrow G^{1/4}HG^{-1/4}$.

Before performing the thin-layer procedure, we need to limit the amplitude of the linear transformation $\mathbb{T}$ since a generally linear transformation prevents the dynamics separation between longitudinal and transverse direction.

Here, we consider the slight variation condition of the cross section, which takes the form $\mathbb{T}=\mathbb{R}_\theta[1+\delta \mathbb{W}]$, where $\mathbb{R}_\theta$ is a 2D rotation matrix and $\mathbb{W}$ is a general $2\times2$ linear transformation matrix. 
Mathematically, we have $|\mathbb{T}|=1+\delta \text{tr}(\mathbb{R}_\theta^{-1} \mathbb{W})+O(\delta^{3/2})$. Furthermore, the variation of the transformation along $s$ should be sufficiently slow, i.e., $\partial_s T_{ab}\ll 1/(\delta q)$.
With this form, we estimate that $\mathcal{A}^a$ is of order $\delta^{1/2}$. Accordingly, the rescaled Hamiltonian can be expanded in power of $\delta$ as
\begin{equation}
G^{1/4}HG^{-1/4}=H_{(-1)}+H_{(0)}+O(\delta^{1/2}),
\end{equation}
where
\begin{equation}\label{h-1}
H_{(-1)}=\frac{1}{\delta}\left[-\frac{\hbar^2}{2m}\nabla_\perp^2+V_c(q_1',q_2')\right],
\end{equation}
with $\nabla_\perp^2=\partial_{q_1'}^2+\partial_{q_2'}^2$,
and
\begin{equation}
\begin{aligned}
H_{(0)}=-\frac{\hbar^2}{2m}[\frac{1}{\sqrt{G}}D_s(\sqrt{G}G^{ss}D_s)]+V_T+V_g+V_H.
\end{aligned}
\end{equation}
Here, the well-known geometric potential $V_g=-\frac{\hbar^2 \kappa^2}{8m}$, and the term $V_H$, which arises from the linear transformation $\mathbb{W}$, has the form
\begin{equation}
V_H=-\frac{\hbar^2}{2m}[2W_{11}\partial_2^2 +2 W_{22}\partial_1^2-2(W_{12}+ W_{21
}) \partial_2\partial_1],
\end{equation}
where $W_{ab}$ are the components of $\mathbb{W}$.

The Hamiltonian $H_{(-1)}$ describes a quantum particle confined in a 2D plane by the potential $V_c(q_1', q_2')$, which governs the eigenstates in the transverse direction. The Hamiltonian $H_{(0)}$, on the other hand, primarily governs the tangential dynamics along the tube. The kinetic energy term $V_H$ originates from the transverse direction and is explicitly induced by the transformation matrix $\mathbb{W}$. Since $H_{(-1)}$ is of order $1/\delta$, the energy levels corresponding to different transverse eigenstates exhibit negligible overlap along the longitudinal direction within the energy range of $H_{(0)}$. This allows us to employ the ansatz $\psi(\bm{q}_\perp',s)=\sum_\beta \chi_\beta(\bm{q}_\perp') \phi_\beta(s)$, where $\beta$ is a degeneracy index and $\chi_\beta$ satisfies the eigenvalue equation $H_{(-1)} \chi_\beta = E_{(-1)} \chi_\beta$. Note that $H_{(0)}$ is not yet the desired effective Hamiltonian, as the operator $D_s$ and the term $V_H$ still depend on transverse coordinates and derivatives.

By projecting onto the subspace spanned by the transverse states $\chi_\beta$, the effective tangential Hamiltonian is obtained as
\begin{equation}\label{heff}
H_{\text{eff}}^{\beta' \beta}=\int \chi_{\beta'}^* H_{(0)} \chi_{\beta} d\bm{q}_\perp'.
\end{equation}
In the following, we derive the explicit forms of this effective Hamiltonian for the cases in which the tube cross section undergoes rotation, scaling, and shear transformations, respectively.

\subsection{Rotation}
For a pure rotational transformation, we have $\mathbb{W} = 0$ and
\begin{equation}
\mathbb{T}=\mathbb{R}_\theta=\left[
\begin{array}{ccc}
\cos\theta(s) & -\sin\theta(s) \\
\sin\theta(s)& \cos\theta(s)
\end{array}
\right].
\end{equation}

In this case, we obtain
\begin{equation}
\bar{\bm{\xi}}=\sqrt{\delta}(\omega+\tau)\left(
\begin{array}{ccc}
q_1'\cos\theta-q_2'\sin\theta \\
q_1'\sin\theta+q_2'\cos\theta
\end{array}
\right),
\end{equation}
where $\omega=\partial_s \theta$.
This leads to
\begin{equation}
\mathcal{A}^1=\sqrt{\delta}q_2'(\omega+\tau )
\end{equation}
and
\begin{equation}
\mathcal{A}^2=-\sqrt{\delta}q_1'(\omega+\tau )
\end{equation}
As a result, $\partial_a \mathcal{A}^a = 0$, which implies $V_T = 0$.
The Hamiltonian $H_0$ therefore reduces to
\begin{equation}
H_0=-\frac{\hbar^2}{2m}D_s^2+V_g,
\end{equation}
where the covariant derivative is given by $D_s=\partial_s-i(\omega+\tau)\hat{L}/\hbar$. 

If the linear transformation is applied to a tube with a circular cross section, the transverse eigenstates can be expressed as $\chi_{n,l} \propto \mathcal{R}_{n,l}(\rho) e^{il\theta}$, where $\rho = \sqrt{q_1'^2 + q_2'^2}$, $l = 0, \pm1, \pm2, \cdots$ is the angular quantum number, $n = 1, 2, 3, \ldots$ denotes the radial quantum number, and $\mathcal{R}_{n,l}(\rho)$ represents the radial wave function. Using Eq.~\eqref{heff}, we obtain the effective Hamiltonian for a circular cross section under a rotational transformation,
\begin{equation}
H_{\text{cir}}=-\frac{\hbar^2}{2m}[\partial_s-i(\omega+\tau)l]^2+V_g.
\end{equation}

If the transformation is applied to a tube with a square cross section, the transverse eigenstates are given by $|n_1n_2\rangle=\frac{2}{d}\sin(\frac{n_1\pi q_1}{d}-\frac{n_1\pi}{2})\sin (\frac{n_2\pi q_2}{d}-\frac{n_2\pi}{2})$, where $d$ is the side length. Due to the square symmetry, the states $|n_1 n_2\rangle$ and $|n_2 n_1\rangle$ are degenerate. A suitable choice of basis states is $|+\rangle_{n_1n_2}=(|n_1n_2\rangle +i |n_2n_1\rangle)/\sqrt{2}$ and $|-\rangle_{n_1n_2}=(|n_1n_2\rangle -i |n_2n_1\rangle)/\sqrt{2}$. The effective Hamiltonian in the subspace spanned by $|+\rangle_{n_1 n_2}$ and $|-\rangle_{n_1 n_2}$ takes the form
\begin{equation}
\begin{aligned}
H_{\text{squ}}= -\frac{\hbar^2}{2m}[\partial_s- i (\omega+\tau)\sigma_z \langle L \rangle_{n_1n_2} ]^2+V_g,
\end{aligned}
\end{equation}
where
$\langle L \rangle_{n_1n_2}=\frac{16n_1^2n_2^2[-1+(-1)^{n_1+n_2}]^2}{(n_1^2-n_2^2)^3 \pi^2}$. It can be observed that when $n_1 + n_2$ is even, the gauge term in the effective Hamiltonian vanishes. In contrast, when $n_1 + n_2$ is odd, the gauge term takes opposite signs for the two basis states.

Therefore, both the rotational transformation and the torsion of the tube itself induce an effective angular momentum that acts as a gauge potential, coupling to the angular momentum eigenvalue. Moreover, in both circular and square cross sections, the rotational transformation does not break the degeneracy of the transverse eigenstates.

\subsection{Scaling}
For a scaling transformation, we have $\theta(s) = 0$, so that $\mathbb{R}_\theta = \mathbf{I}$, where $\mathbf{I}$ is the $2 \times 2$ identity matrix, and $\mathbb{W}$ is given by
\begin{equation}
\mathbb{W}=\left[
\begin{array}{ccc}
f_1(s) & 0 \\
0 & f_2(s)
\end{array}
\right].
\end{equation}
If the scaling corresponds to a stretching transformation, one of the functions $f_1$ or $f_2$ is set to unity. In the case of a squeezing transformation, the condition $(1 + \delta f_1)(1 + \delta f_2) = 1$ must be satisfied, implying $f_1 = -f_2$.

In this case,
\begin{equation}
\bar{\bm{\xi}}=\sqrt{\delta}\left[
\begin{array}{ccc}
(1+\delta \dot{f}_1)\tau q_1'+\delta \dot{f}_2 q_2' \\
-\delta \dot{f}_1 q_1'+(1+\delta \dot{f}_2)\tau q_2'
\end{array}
\right].
\end{equation}
Consequently, we obtain
\begin{equation}
\mathcal{A}^1=(1+\delta f_2)\sqrt{\delta}[-\delta \dot{f}_1 q_1'+(1+\delta \dot{f}_2)\tau q_2']
\end{equation}
and
\begin{equation}
\mathcal{A}^2=-(1+\delta f_1)\sqrt{\delta}[(1+\delta \dot{f}_1)\tau q_1'+\delta \dot{f}_2 q_2']
\end{equation}
One can find that $V_T\sim O(\delta)$ and $\partial_a \mathcal{A}^a \sim O(\delta)$.
Therefore, to zeroth order in $\delta$, the covariant derivative remains $D_s=\partial_s-i\tau \hat{L}/\hbar$. When evaluating $V_H$, different cases must be treated separately.

When the tube has a circular cross section, the effective Hamiltonian in the subspace spanned by $\chi_{n,l}$ and $\chi_{n,-l}$ is given by
\begin{equation}
\begin{aligned}
V_H^{ll'}=\delta(f_1+f_2)E_{n,l}\textbf{I},
\end{aligned}
\end{equation}
where $E_{n,l}$ denotes the transverse eigenenergy of the degenerate states. The resulting effective Hamiltonian becomes
\begin{equation}
H_{\text{cir}}=-\frac{\hbar^2}{2m}(\partial_s-i\tau l)^2+V_g+\delta(f_1+f_2)E_{n,l}.
\end{equation}

For a tube with a square cross section, the correction term within the subspace spanned by $|+\rangle_{n_1n_2}$ and $|-\rangle_{n_1n_2}$ takes the form
\begin{equation}
\begin{aligned}
V_{H(\text{scaling})}^{n_1n_2}=&\delta(f_1+f_2)(E_{n_1}+E_{n_2})\textbf{I} \\
&+\delta(f_1-f_2)(E_{n_2}-E_{n_1})\sigma_x,
\end{aligned}
\end{equation}
where $E_{n_a}$ is the transverse eigenenergy along the $q_a'$ direction. The corresponding effective Hamiltonian is then
\begin{equation}
H_{\text{squ}}=-\frac{\hbar^2}{2m}(\partial_s-i\tau \sigma_z \langle L\rangle_{n_1n_2})^2+V_g+V_{H(\text{scaling})}^{n_1n_2}.
\end{equation}

It can be observed that for a tube with a circular cross section, slight scaling introduces an effective potential without lifting the degeneracy of the energy levels. In contrast, for a square cross section, the effective potential resulting from scaling contains non-zero off-diagonal elements in excited modes where $n_1\neq n_2$, leading to a splitting of the degenerate levels.
Notably, under a squeezing transformation where $f_1=-f_2$, the correction $V_H$ vanishes in the circular case, while in the square case only the off-diagonal components remain. This suggests that waveguides with circular cross sections are less susceptible to deformation-induced mode mixing than those with square cross sections, thereby offering better preservation of wave modes.


\subsection{Shearing}
For a shearing transformation parallel to the $q_1$ axis, we have $\theta(s) = 0$ and
\begin{equation}
\mathbb{W}=\left(
\begin{array}{ccc}
0 & f(s) \\
0 & 0
\end{array}
\right)
\end{equation}
One finds that
\begin{equation}
\bar{\bm{\xi}}=\sqrt{\delta}\left[
\begin{array}{ccc}
\tau (q_1'+\delta f q_2') \\
(-\delta \dot{f} +\tau) q_2'
\end{array}
\right].
\end{equation}
Subsequently, we obtain
\begin{equation}
\mathcal{A}^1=\sqrt{\delta}[\delta f \tau(q_1'+\delta f q_2')+(-\delta \dot{f}+\tau)q_2']
\end{equation}
and
\begin{equation}
\mathcal{A}^2=-\tau\sqrt{\delta}(q_1'+\delta f q_2')
\end{equation}
Similar to the scaling case, here $\partial_a \mathcal{A}^a \sim O(\delta)$ and $V_T\sim O(\delta)$. The covariant derivative remains $D_s=\partial_s-i\tau \hat{L}/\hbar$, and and the resulting term $V_H$ is given by
$V_H=\frac{\hbar^2}{m}f \partial_1\partial_2$.

Performing the integral within the subspace of degenerate states for the circular and square cross section cases yields  $V_H^{ll'}=0$ and
\begin{equation}
V_{H(\text{shear})}^{n_1n_2}=-\frac{32f}{\pi^2 } \frac{n_1^2 n_2^2 (E_{n_1}+E_{n_2})}{(n_1^2-n_2^2)^2(n_1^2+n_2^2)} \sigma_x ,
\end{equation}
respectively.
Thus, the effective Hamiltonians for the circular and square cross-sections in this case are respectively
\begin{equation}
H_{\text{cir}}=-\frac{\hbar^2}{2m}(\partial_s-i\tau l)^2+V_g,
\end{equation}
and
\begin{equation}
H_{\text{squ}}=-\frac{\hbar^2}{2m}(\partial_s-i\tau \sigma_z \langle L\rangle_{n_1n_2})^2+V_g+V_{H(\text{shear})}^{n_1n_2}.
\end{equation}
Once again, the difference between the Hamiltonians for the circular and square tubes under shearing transformation demonstrates that circular waveguides are more effective in preserving propagation modes.

\section{Combination of rotation and squeezing in helical tubes: state phase and quantum geometric tensor }\label{sec4}
In this section, we consider a concrete example in which a tube with a square cross section undergoes a linear transformation that combines rotation with squeezing. The transformation matrix is given by $\mathbb{T} = \mathbb{R}\theta[1 + \delta \mathbb{W}]$, where
\begin{equation}
\mathbb{W}=\left[
\begin{array}{ccc}
f(s) & 0 \\
0 & -f(s)
\end{array}
\right].
\end{equation}
Then, in the subspace spanned by the basis states $|+\rangle_{n_1n_2}$ and $|-\rangle_{n_1n_2}$, the effective equation becomes
\begin{equation}\label{eqs}
-\frac{\hbar^2}{2m}[\partial_s-i\alpha\sigma_z ]^2\phi(s)+V_{H(\text{squeeze})}^{n1n2}\phi(s)=E\phi(s),
\end{equation}
where $\alpha=(\omega+\tau) \langle L \rangle_{n_1n_2}$ and the potential $V_{H(\text{squeeze})}^{n1n2}$ in this basis is
\begin{equation}
V_{H(\text{squeeze})}^{n1n2}=\left[
\begin{array}{ccc}
0 & 2f\Delta E \\
2f\Delta E & 0
\end{array}
\right],
\end{equation}
wherein $\Delta E=E_{n_2}-E_{n_1}$.

Obtaining an exact analytical solution to this equation is challenging because the gauge potential is diagonal while the effective potential is off-diagonal. Here, we assume that $E\gg |2f\Delta E|$, $E\gg |\hbar^2 \alpha^2/(2m)|$, and that the functions $\tau(s)$, $f(s)$ and $\omega(s)$ vary slowly. Under these conditions, we can apply the WKB approximation by treating the effective Hamiltonian in a classical form and neglecting the term $\partial_s(\omega+\tau)$. This leads to two energy branches
\begin{equation}
\mathcal{E}_{\pm}=\frac{p^2}{2m}+\frac{\hbar^2\alpha^2}{2m}\pm \sqrt{(\frac{\hbar \alpha p}{m})^2+(2f\Delta E)^2},
\end{equation}
where $p$ denotes the classical momentum.
Setting $E=\mathcal{E}_\pm$, we solve for the momentum
\begin{equation}
p_\pm \approx \pm\hbar \alpha \pm \sqrt{2m(E\mp 2f\Delta E)-\hbar^2\alpha^2}.
\end{equation}
In deriving this expression for the momentum, the coupling term between $p_+$ and $p_-$ has been neglected. Notably, motion in opposite directions corresponds to distinct energy levels. This indicates that geometric chirality formed by rotation and squeezing transformations induces a splitting of the energy levels, thereby lifting the degeneracy of previously degenerate states and coupling the direction of motion to specific energy levels. This phenomenon is reminiscent of spin-momentum locking in topological insulators.

For each energy branch, we assume a WKB solution of the form
\begin{equation}
\phi_{\pm}=\bm{u}_\pm(s) \exp(\frac{i}{\hbar}\int p_{\pm} ds),
\end{equation}
where $\bm{u}_\pm(s)$ can be interpreted as slowly varying SO(2) spinors.

Substituting the WKB ansatz into Eq.~\eqref{eqs} and retaining terms to zeroth order in $\hbar$, we obtain
\begin{equation}
\bm{u}_+=\frac{C_+}{\sqrt{|p_+-\hbar\alpha|}}\frac{1}{\sqrt{2\Lambda_+(\Lambda_+-\frac{\hbar \alpha p_+}{m})}}\left[
\begin{array}{ccc}
2f\Delta E \\ \Lambda_+-\frac{\hbar \alpha p_+}{m}
\end{array}
\right],
\end{equation}
and
\begin{equation}
\bm{u}_-=\frac{C_-}{\sqrt{|p_-+\hbar\alpha|}}\frac{1}{\sqrt{2\Lambda_-(\Lambda_- +\frac{\hbar \alpha p_-}{m})}}\left[
\begin{array}{ccc}
-\Lambda_- -\frac{\hbar \alpha p_-}{m} \\ 2f\Delta E
\end{array}
\right],
\end{equation}
where $\Lambda_\pm=\sqrt{(2f\Delta E)^2+(\frac{\hbar \alpha p_\pm}{m})^2}$ and $C_\pm$ are normalization constants.

\begin{figure}
  \centering
  \includegraphics[width=0.45\textwidth]{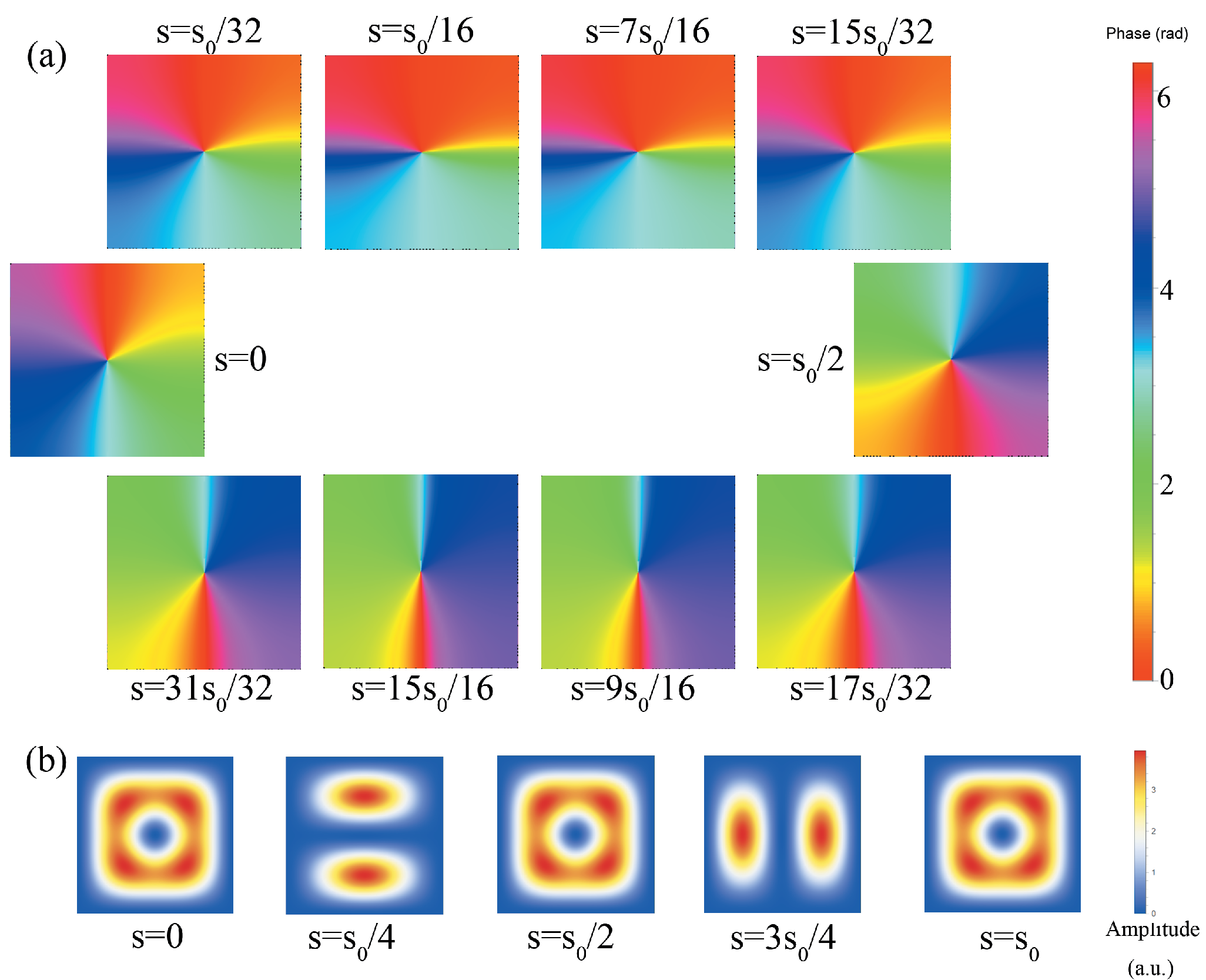}\\
  \caption{Evolution of the state $\psi_+$ in the tube's cross section for $n_1=1$ and $n_2=2$. (a) Phase evolution. (b) Amplitude evolution.}\label{fig2}
\end{figure}

As we have assumed $E\gg \hbar^2 \alpha^2/(2m)$, which implies $p_\pm\gg \hbar \alpha$, the difference between $|p_+|$ and $|p_-|$ can be neglected. The primary distinction between $\phi_+$ and $\phi_-$ thus lies in their spinor components $\bm{u}_{\pm}$.
The evolution of $\bm{u}_{\pm}$ can be visualized through the phase and amplitude of $\psi_{\pm}=\bm{u}_{\pm}\cdot [(|+\rangle_{n_1n_2},0)^\mathrm{T}+(0,|-\rangle_{n_1n_2})^\mathrm{T}]$ within the cross section.
We choose $\delta=0.02$, $\omega=\tau=0.02/a_0$, where $a_0$ is the characteristic size of the cross section, and $f=\sin (2\pi s/s_0)$, with $s_0$ denoting the total length of the tube. Within the degenerate subspace corresponding to $n_1=1$ and $n_2=2$, we plot the phase and amplitude evolution of $\psi_+$ in Fig.~\ref{fig2}.

At $s=0$, where the potential vanishes, the phase profile of $|+\rangle_{12}$ is trisected symmetrically (Fig.~\ref{fig2}(a)), and the amplitude distribution exhibits square symmetry (Fig.~\ref{fig2}(b)), reflecting the degeneracy of the energy levels. As the propagation distance $s$  increases, the phase distribution gradually transitions from a trisected to a bisected pattern. Between $s=s_0/16$ and $s=7s_0/16$,  the phase remains stable, and the amplitude distribution shifts from square to a bifurcated form. Approaching $s=s_0/2$, as the influence of the potential diminishes, the phase begins to revert to a trisected distribution, and the amplitude recovers its square symmetry. Beyond $s=s_0/2$, the phase exhibits a $\pi$jump, and the amplitude abruptly transitions again to a bifurcated state, though now rotated by$\pi/2$ compared to the configuration observed at $s=s_0/4$. From $s=s_0/2$ to $s=s_0$, the phase and amplitude evolve in a manner similar to the first half of the propagation. A comparison between the expressions for $\bm{u}+$ and $\bm{u}-$ reveals that the phase associated with $\bm{u}-$ differs from that of $\bm{u}+$ by approximately $\pi/2$, while their evolutionary behaviors remain similar.

Treating $f$ and $\omega$ as slowly varying parameters and considering the Hamiltonian as a two-level system, we can approximately derive the quantum geometric tensor $Q_{\mu \nu}$ in the parameter space $(\omega,f)$,
\begin{equation}
Q_{\mu\nu}=g_{\mu\nu}-\frac{i}{2}F_{\mu\nu},
\end{equation}
where the Riemannian metric is given by
\begin{equation}
g_{\mu\nu}=\frac{1}{\Lambda^2}\left[
\begin{array}{ccc}
B^2(p^2 \sin^2 2\varphi+\hbar^2 \dot{\varphi}^2 \cos^2 2\varphi) & -\frac{Bp\Delta E}{2}\sin 4\varphi \\ -\frac{Bp\Delta E}{2}\sin 4\varphi & \Delta E^2 \cos^2 2\varphi
\end{array}
\right],
\end{equation}
and the Berry curvature is
\begin{equation}
F_{\mu\nu}=\frac{1}{\Lambda^2}\left[
\begin{array}{ccc}
0 & -2\Delta E B\hbar\dot{\varphi}\cos^2  2\varphi \\ 2\Delta E B\hbar\dot{\varphi}\cos^2  2\varphi & 0
\end{array}
\right],
\end{equation}
with $B=\frac{\hbar}{2m}\langle L\rangle_{n_1n_2}$. In this derivation, we have used the approximations $|p_+|\approx |p_-|=p$ and $\Lambda_+ \approx \Lambda_-=\Lambda$. We have also define $\cos\varphi=\frac{2f\Delta E}{\sqrt{2\Lambda(\Lambda-\frac{\hbar \alpha p}{m})}}$ and $\sin\varphi=\frac{\Lambda-\frac{\hbar \alpha p}{m}}{\sqrt{2\Lambda(\Lambda-\frac{\hbar \alpha p}{m})}}$.

It is noted that the metric $g_{\mu\nu}$ may exhibit singular behavior when $\varphi$ approaches 0 or $\pi/2$, which occurs as the quantum state becomes degenerate. The presence of off-diagonal elements indicates a non-trivial coupling between the parameters $\omega$ and $f$. The existence of the Berry curvature implies that during an adiabatic cyclic evolution in parameter space, the system accumulates a geometric phase, whose value is related to the integral of the curvature over the enclosed area.

\section{Conclusion}\label{sec5}
In this work, we have systematically investigated the effective quantum dynamics of a particle confined in a twisted tube with a cross section subject to linear transformations along the axial direction. By extending the thin-layer procedure to adapted curvilinear coordinates, we derived a general effective Hamiltonian that incorporates both geometric effects and contributions induced by the slight transformations.

Our analysis demonstrates that three different linear transformations, namely rotation, scaling, and shearing, distinctly shape the effective quantum behavior. Rotation introduces a geometric gauge field coupled to angular momentum, while scaling and shearing produce additional potential terms. Notably, square cross sections exhibit transformation induced energy level splitting, in contrast to circular ones, which remain invariant under such mode-mixing effects.
Through a concrete example involving simultaneous rotation and squeezing, we further demonstrated how geometric transformations can lead to chirality-dependent energy branches and non-trivial phase evolution in degenerate subspaces. The computed quantum geometric tensor offers deeper insight into the parameter sensitivity and Berry curvature effects in such constrained systems.

These findings not only advance our understanding of deformation induced quantum phenomena in low-dimensional and curved structures but also suggest practical strategies for designing nanoscale waveguides with tailored spectral and transport properties. The methodology and conclusions presented in this work are also applicable to other one-dimensional scalar wave systems, such as scalar optical waveguides and acoustic wave systems.

\acknowledgments

This work is supported in part by the National Natural Science Foundation of China (under Grants No. 12104239), the Natural Science Foundation of Jiangsu Province (under Grant No. BK20210581). M. Y. L is supported by the Jiangxi Provincial Natural Science Foundation with Grant No. 20224BAB211020.

\bibliographystyle{apsrev4-1}
\bibliography{ref19}

\end{document}